\documentclass[a4paper]{IEEEtran}
\usepackage[a4paper, total={7in, 9.8in}]{geometry}
\usepackage[boxruled]{algorithm2e}
\usepackage{enumerate}
\usepackage{multirow,array}
\usepackage{cite}
\usepackage{graphicx}
\usepackage{psfrag}
\usepackage{caption}
\usepackage{subcaption}
\usepackage{url}
\usepackage{amsmath}
\usepackage{amsthm}
\usepackage{array}
\usepackage{algorithm2e}
\usepackage{amssymb}
\usepackage{amsfonts}
\usepackage{float}
\makeatletter
\renewcommand{\boxed}[1]{\text{\fboxsep=.2em\fbox{\m@th$\displaystyle#1$}}}
\makeatother
\newcommand{\C}{\cal}
\newcommand{\Lbarsq}{\overline{{\cal L}^2}}

\newcommand{\fq}{{\mathbb F}_q}
\newcommand{\gbinom}[2]{\begin{bmatrix}#1\\#2\end{bmatrix}_q}

\newtheorem*{conjecture*}{Conjecture}
\newtheorem{lemma}{Lemma}

\newtheorem{definition}{Definition}
\newtheorem{theorem}{Theorem}

\title{Coded Caching via Projective Geometry: ~\\A new low subpacketization scheme}
\begin{document}

\author{
\IEEEauthorblockN{Hari Hara Suthan C, Bhavana M, Prasad Krishnan\\}
\IEEEauthorblockA{
Signal Processing and Communications Research Center,\\
International Institute of Information Technology, Hyderabad.\\
Email: \{hari.hara@research., bhavana.mvn@research., prasad.krishnan@\}iiit.ac.in}
}
\date{\today}
\maketitle
\thispagestyle{empty}	
\pagestyle{empty}
\begin{abstract}
Coded Caching is a promising solution to reduce the peak traffic in broadcast networks by prefetching the popular content close to end users and using coded transmissions. One of the chief issues of most coded caching schemes in literature is the issue of large \textit{subpacketization}, i.e., they require each file to be divided into a large number of subfiles. In this work, we present a coded caching scheme using line graphs of bipartite graphs in conjunction with projective geometries over finite fields. The presented scheme achieves a rate $\Theta(\frac{K}{\log_q{K}})$ ($K$ being the number of users, $q$ is some prime power) with  \textit{subexponential} subpacketization $q^{O((\log_q{K})^2)}$ when cached fraction is upper bounded by a constant ($\frac{M}{N}\leq \frac{1}{q^\alpha}$) for some positive integer $\alpha$). Compared to earlier schemes, the presented scheme has a lower subpacketization (albeit possessing a higher rate). We also present a new subpacketization dependent lower bound on the rate for caching schemes in which each subfile is cached in the same number of users. Compared to the previously known bounds, this bound seems to perform better for a range of parameters of the caching system.  
\end{abstract}


\section{Introduction}
The key performance challenges that next generation wireless networks (5G) face are low latency, high throughput and energy efficiency\cite{FTB}. Content delivery networks have been estimated to carry $72\%$ of the global internet traffic by $2022$ \cite{Cis}. 
\textit{Coded caching} was proposed recently in a landmark paper by Maddah-Ali and Niesen \cite{MaN} and has emerged as an important tool to address major challenges of future communication networks. 
Since its inception coded caching has proved as an efficient tool to trade-off expensive bandwidth with abundantly available and cost-effective memory 
at the user/network nodes.

In \cite{MaN}, the setup consists of a single server with $N$ equi-popular files of same size (divided into $F$ subfiles each of the same size, where $F$ is known as the \textit{subpacketization} parameter), and $K$ users(clients) each having a local memory called \textit{cache} that can store $MF$ subfiles. The \textit{centralized coded caching} scheme of \cite{MaN} works in two phases. In the \textit{caching phase} (which occurs during off peak times) the cache of each client is populated with some $\frac{M}{N}$ fraction of each file in the server.
In the \textit{delivery phase} (which happens during peak traffic times), the clients demand one file each from the server, to satisfy which the server sends coded transmissions. 
The \textit{rate} $(R)$ of such a coded caching scheme is defined as the ratio of the number of bits transmitted to the size of each file, which can be calculated as
\[
\text{Rate}~R = \small \frac{\text{Number of transmissions in the delivery phase}}{\text{Number of subfiles in a file}},
\] when each transmission is of the same size as the subfiles. The Ali-Niesen scheme in \cite{MaN} achieves $R=\frac{K(1-\frac{M}{N})}{\gamma}$, where $\gamma=1+\frac{KM}{N}$ is the \textit{global caching gain}, i.e., the number of users served by each transmission in the delivery scheme. This rate was shown to be optimal for uncoded cache placement \cite{WTP}. Further,  the subpacketization level used by the Ali-Niesen scheme to achieve this rate is $F=\binom{K}{\frac{KM}{N}}$.  Note that as $K$ grows large, $F\approx 2^{KH(M/N)}$, (for constant $\frac{M}{N}$, $H(.)$ being the binary entropy). This means that the files have to be extremely large  for even $50$-$100$ clients, making the Ali-Niesen scheme impractical for applications. 

Since then several new coded caching schemes with lower subpacketization have been constructed at the cost of increase in rate, or cache requirement, or the number of users \cite{TaR,YCTCPDA,cheng2017coded,YTCC}. Among these, an important construction was reported in \cite{YCTCPDA}, via a combinatorial object that the authors defined, known as \textit{Placement Delivery Arrays} (PDAs). The PDA constructed in \cite{YCTCPDA} achieved a global caching gain of $\frac{MK}{N}$ (one less than that of \cite{MaN}), while improving the subpacketization by an exponential factor compared to \cite{MaN}. However, in this construction, as well as in most others in literature, the subpacketization required for the caching schemes continues to be exponential in $K^{\frac{1}{r}}$ (for some positive integer $r$) to the best of our knowledge.  

Recently, a line graph based approach to coded caching was introduced in \cite{PK}. 
Using this framework, a construction for a caching scheme was given via a projective geometry over a finite field. The scheme presented in \cite{PK} achieves a constant rate with subpacketization subexponential in $K$ $\left(F=q^{O((log_q K)^2)}\right)$ for some prime power $q$). However the drawback of this scheme is that the uncached fraction of each file has to be large $\left((1-\frac{M}{N})=\Theta(\frac{1}{\sqrt{K}})\right)$. We remedy this drawback (to some extent) in this work. 


The contributions and organization of this paper are as follows. In Section \ref{review}, we review the line graph based coded caching scheme proposed in \cite{PK}, while refining it slightly for our purposes. In Section \ref{lowerbound}, we propose a new lower bound for the optimal rate $R^*$ given parameters $K, F,$ and $\frac{M}{N}$, for the caching schemes in which each subfile is cached in the same number of users, which is a property satisfied by all known centralized caching designs in literature (to the best of our knowledge). Using some numerical examples, we see that this lower bound performs better (for a range of parameters) compared to the previously known bounds in \cite{WTP,cheng2017coded}. In Section \ref{ourscheme}, we present a new coded caching scheme using projective geometries over finite fields in the line graph framework of \cite{PK}. In Section \ref{asymptotics} we give the asymptotic analysis of the scheme proposed. We show that the scheme achieves a rate $R=\Theta\left(\frac{K}{log_q K}\right)$ ($q$ being a prime power) for a constant cache requirement, which can be extremely small ($\frac{M}{N}\leq \frac{1}{q^{\alpha}}$, for some constant positive integer $\alpha$). The subpacketization achieved is subexponential in $K$, $F=q^{O((\log_q{K})^2)}$. We provide a table in Section \ref{asymptotics} which compares the parameters of our scheme to that of \cite{YCTCPDA}.  

\textit{Notations and Terminology:} $\mathbb{Z}^{+}$ denotes the set of positive integers.
We denote the set $\{1,\hdots,n\}$ by $[n]$ where $n\in \mathbb{Z}^{+}$.
We give the basic definitions in graph theory. The sets $V(G), E(G)$ denote vertex set and edge set of a graph $G$ respectively, where $E(G)\subseteq \left\{\{u,v\}:u,v\in V(G)\right\}$. All graphs considered in this paper are undirected graphs with no self loops. The neighbourhood of a vertex $u\in V(G)$ is given as $\mathcal{N}(u)=\{v\in V(G): \{u,v\}\in E(G)\}$. The square of a graph $G$ is a graph $G^2$ having $V(G^2)=V(G)$ and an edge $\{u,v\}\in E(G^2)$ if and only if either $\{u,v\}\in E(G)$ or there exists some $v_1\in V(G)$ such that $\{u,v_1\},\{v_1,v\}\in E(G)$. The complement of a graph $G$ is denoted as $\overline{G}$. A set $H\subseteq V(G)$ is called a clique of $G$ if every two distinct vertices in $H$ are adjacent to each other. A single vertex is also considered as a clique by definition. A clique cover of $G$ is a collection of disjoint cliques such that each vertex appears in precisely one clique. 
A \textit{bipartite graph} $B$ is a graph whose vertices can be partitioned into two independent sets (called left and right vertices of $B$) such that edges exist only between left and right vertices. 
A bipartite graph is (left or right) regular if the degree of each vertex on (left or right) is same throughout (left or right) partition. A bipartite graph is bi-regular if it is both left regular and right regular. For more information on graph theory the reader is referred to \cite{Die}.

\section{Review of Line Graph scheme in \cite{PK}}
\label{review}
We now review the basic framework and some results of \cite{PK}. Consider a coded caching system consisting of a server with files $\{W_i:i\in [N]\}$. Let $\mathcal{K}$ be any set such that $|\mathcal{K}|=K$. We shall use $\mathcal{K}$ to indicate the set of $K$ users. Let $\mathcal{F}$ be any set such that $|\mathcal{F}|=F$. The subfiles of a file $W_i$ are denoted by $W_{i,f}$ where $f\in \mathcal{F}$ and $W_{i,f}$ takes values in some Abelian group. Here we consider \textit{symmetric caching}, i.e., for any $f\in \mathcal{F}$, either a user caches $W_{i,f}, \forall i\in[N]$ or the user does not cache $W_{i,f}$  for any $i\in[N]$. Any symmetric caching scheme can be represented as an equivalent $D$-left regular \textit{bipartite caching graph} $B(K,D,F)$ with left vertices being $\mathcal{K}$ and right vertices being $\mathcal{F}$, and the uncached fraction $1-\frac{M}{N}=\frac{D}{F}$. The uncached subfiles are identified by the edges of $B$ i.e, for $k\in\mathcal{K}, f\in\mathcal{F}$ an edge $\{k,f\}\in E(B)$ if and only if the subfiles $W_{i,f}, \forall i\in[N]$ are not present in the cache of user $k$. This bipartite coded caching setup was given in \cite{YTCC}. In \cite{PK}, a line graph based framework was proposed  to study the coded caching problem. The \textit{line graph} ${\C L}(G)$ (or simply, ${\cal L}$) of an undirected graph $G$ is a graph in which the vertex set $V({\C L}(G))$ is the edge set $E(G)$ of $G$, and two vertices of $V({\C L}(G))$ are adjacent if and only if they share a common vertex in $G$. The caching scheme was captured via a line graph ${\mathcal{L}}$ of a bipartite caching graph and the delivery scheme was obtained as a clique cover of complement of the square of line graph denoted by ${\Lbarsq}$. The following lemma proved in \cite{PK} presents the conditions under which an arbitrary graph is a line graph of a left regular bipartite graph. This enables us to construct a line graph which corresponds to a coded caching scheme.

\begin{lemma} \cite{PK}
\label{linegraph}
A graph ${\C L}$ containing $KD$ vertices is the line graph of a $D$-left-regular bipartite graph $B(K,D,F)$ if and only if the following conditions are satisfied.
\begin{itemize}
\item The vertices of $\C L$ can be partitioned into $K$ disjoint cliques containing $D$ vertices each. We denote these cliques by ${\C U}_{k}:k\in \mathcal{K}$ and call them as the \underline{user-cliques}. 
\item Consider distinct $k_1,k_2\in\mathcal{K}.$ For any vertex $v\in {\C U}_{k_1}$, there exists at most one vertex $w\in {\C U}_{k_2}$ such that $\{v,w\}\in E({\C L}).$ 
\item For any $k \in {\C K}$ and any vertex $v\in {\C U}_{k}$,  the set $\{v\}\cup  {\C N}(v)\setminus {\C U}_{k}$ (containing $v$ and all adjacent vertices of $v$ except those in ${\C U}_{k}$), forms a clique. We refer to these cliques as the \underline{subfile-cliques}.  
\item Let $r$ be the number of subfile-cliques in $\C L$ and the subfile-cliques be denoted as ${\C S}_i:i\in [r]$. Then the number of right vertices of $B$ is $F=r$.

\label{subpacket}
\end{itemize}

\end{lemma}

Any graph $\cal {L}$ that satisfies the above conditions for some $K$ and $D$ is called as a \textit{caching line graph}. Since there are $r=F$ subfile cliques, we can denote the subfile-cliques as  ${\C S}_f:f\in {\C F}$. It also holds by the construction of ${\C L}$ that there is at most one vertex in the intersection of any  given user-clique ${\C U}_k$ and a subfile-clique ${\C S}_f$. Further, note that the subfile-cliques also partition the vertices of ${\C L}$. Thus each vertex of ${\C L}$ lies precisely in  one user-clique and one subfile-clique. Therefore the vertices of ${\C L}$ can be indexed using a subset of ${\C K} \times {\C F}$, i.e.,  $V({\cal L})=\{(k,f)\in{\cal K}\times{\cal F}:{\cal U}_k\cap{\cal S}_f\neq \phi\}$. With this notation, we have ${\C U}_k=\{(k,f) \in V({\C L}):f\in {\C F}\}$ and ${\C S}_f=\{(k,f) \in V({\C L}):k\in {\C K}\}$. Furthermore, it follows that $E({\cal L})=\{\{(k,f),(k',f')\}\subset V({\cal L}): k=k' \text{ or } f=f' \text{ but not both }\}$.


Following \cite{PK}, the delivery scheme follows according to a clique cover of $\Lbarsq$. In order to find the cliques in ${\Lbarsq}$ (called \textit{transmission cliques}), first we need to identify the structure of $ {\Lbarsq}$. It is easy to see that $V({\Lbarsq})=V({\C L})$. In Lemma \ref{edge in compliment of square} we present the conditions under which an edge exist in ${\Lbarsq}$. We will use this lemma in Section \ref{ourscheme} to identify such \textit{transmission cliques} in the construction we give.

\begin{lemma}\label{edge in compliment of square}
Let  $(k_1,f_1),(k_2,f_2) \in V(\mathcal{{\C L}})$.
The edge $\{(k_1,f_1),(k_2,f_2)\}\in E({\Lbarsq})$ 
if and only if  $k_1\neq k_2, f_1\neq f_2$ and $(k_1,f_2), (k_2,f_1)\notin V(\mathcal{L})$.
\end{lemma}

\begin{IEEEproof}
The If part of the lemma follows from the definition of $\Lbarsq$. We prove the only if part here. Let $\{(k_1,f_1),(k_2,f_2)\}\in E({\Lbarsq})$. Suppose $k_1=k_2$. Then by the construction of $\mathcal{L}$ we have $\{(k_1,f_1),(k_2,f_2)\}\in E({\C L})$. Therefore $\{(k_1,f_1),(k_2,f_2)\}\notin E({\Lbarsq})$ which is a contradiction. Hence $k_1\neq k_2$. Similarly $f_1\neq f_2$. Suppose $(k_1,f_2) \in V(\mathcal{L})$. Since $(k_1,f_1),(k_2,f_2)\in V(\mathcal{L})$. By the construction of ${\C L}$ we have $\{(k_1,f_2),(k_1,f_1)\}\in E(\mathcal{L})$ and $\{(k_1,f_2),(k_2,f_2)\}\in E(\mathcal{L})$. By definition of $\mathcal{L}^2$, we have $\{(k_1,f_1),(k_2,f_2)\}\in E(\mathcal{L}^2)$ which is a contradiction. Hence $(k_1,f_2) \notin V({\C L})$. Similarly $(k_2,f_1) \notin V({\C L})$.
\end{IEEEproof}

In \cite{PK} a particular class of caching line graphs called $(c,d)$-\textit{caching line graphs} was considered. The advantage of these line graphs is that the corresponding caching scheme parameters are obtained naturally in a simple fashion. This is captured in the following definition and theorem. Our construction in Section \ref{ourscheme} is also based on such $(c,d)$-caching line graphs.
\begin{definition}\cite{PK}
A caching line graph $\C L$ such that $\C L$ has a clique cover consisting of $c$-sized disjoint subfile cliques and $\Lbarsq$ has a clique cover consisting of $d$-sized disjoint cliques, for some $c,d \in \mathbb{Z}^+$, is called a $(c,d)$-caching line graph.
\end{definition}
%
\begin{theorem}\cite{PK}
\label{cliquecoverlinegraph}
Consider a $(c,d)$-caching line graph ${\C L}$. Then there is a coded caching scheme consisting of the caching scheme given by ${\C L}$ with $F=\frac{KD}{c}$ (and thus $\frac{M}{N}=1-\frac{c}{K}$), and there is an associated   transmission scheme based on the clique cover of $\Lbarsq$ having rate $R=\frac{c}{d}$. 
\end{theorem}

\section{A new lower bound on the rate}
\label{lowerbound}

In this section, we propose a lower bound on rate of the delivery scheme for symmetric caching schemes where each subfile is stored in equal number of users. Most known schemes in literature satisfy this property to best of our knowledge. From Section \ref{review}, we know that any symmetric caching scheme can be represented by a left-regular bipartite caching graph $B(K,D,F)$ where $1-\frac{M}{N}= \frac{D}{F}$. 
As each subfile is not cached at equal number of users, the equivalent bipartite graph with $KD$ edges will be right regular as well, with right degree being $K(1-\frac{M}{N})$ for every subfile in $B$.

We first recall a generic lower bound given in \cite{PK} based on structure of $B(K,D,F)$ which is used in Theorem \ref{LB} to arrive at our new bound. Let $H$ be the subgraph of $B$ induced by the vertices $\mathcal{K'}\cup \mathcal{F'}$ where $\mathcal{K'}\subseteq \mathcal{K}$ and $\mathcal{F'}\subseteq \mathcal{F}$. Let $N'=min(|\mathcal{K'}|, K(1-\frac{M}{N})).$ Let $U=\{k_j:j\in[N']\}$ be a subset of $N'$ vertices of $\mathcal{K'}$ taken in some order. For $j\in[N']$, let $\rho_j$ be the set of right vertices (subfiles) in $H$ which are adjacent to $\{k_i:i\in[j]\}.$
Let $R^{*}$ be the infimum of all achievable rates for coded caching problem defined by $B$. Then, from Theorem 2 of \cite{PK},

\begin{equation}\label{eqn1} 
\small
R^*F\geq \sum\limits_{j=1}^{N'}\rho_j.
\end{equation}


We now obtain a lower bound for the case of symmetric caching schemes defined by biregular bipartite caching graphs. The proof for this bound is based on a similar `nested' bound shown in \cite{cheng2017coded}. 
\begin{theorem} \label{LB}
Let $R^{*}$ be the infimum of all achievable rates for the coded caching problem defined by a bi-regular bipartite graph $B$. Then
\small
\begin{align*} \nonumber
R^*F &\geq D+ \left\lceil{\tiny\frac{D(K(1-\frac{M}{N})-1)}{K-1}}\right\rceil +  
\cdots\\ &
\hspace{-0.9cm}\cdots +
\left\lceil{\frac{1}{\frac{KM}{N}+1}\left\lceil {\frac{2}{\frac{KM}{N}+2}\left\lceil\cdots\left\lceil{\frac{D(K(1-\frac{M}{N})-1)}{K-1}}\right\rceil\cdots\right\rceil}\right\rceil}\right\rceil.
\end{align*}
\end{theorem}

\begin{IEEEproof}
Every user vertex has degree $D= F(1-\frac{M}{N})$ in $B$. Consider a user vertex in $B$ and call it as $k_1$. So, by notations of (\ref{eqn1}), $|\rho_1|= D$. Consider the graph induced by $\mathcal{K} \cup {\cal N}(k_1)$ vertices of $B$. Call it $G'$. Since $B$ is a bi-regular graph, degree of each subfile vertex $f \in \mathcal{F} \in B$ is exactly $K(1-\frac{M}{N})$. So, by pigeon-holing argument, it is not difficult to see that there exists a user with degree at least
$\left\lceil{\frac{(K(1-\frac{M}{N})-1)D}{K-1}}\right\rceil$
in $G'$. Consider such user vertex and call it as $k_2$. Then 
$|\rho_2| \geq \left\lceil{\frac{(K(1-\frac{M}{N})-1)D}{K-1}}\right\rceil$. In general, for each $j \in \{2,\cdots,K(1-\frac{M}{N})-1\}$, for the graph induced by $\mathcal{K}\cup {\cal N}(k_1) \cup {\cal N}(k_2)\cdots \cup {\cal N}(k_{j-1})$ vertices of $B$, by pigeon-holing argument there exists a user with degree at least $\left\lceil{\frac{(K(1-\frac{M}{N})-(j-1)}{K-(j-1)}\left\lceil {\cdots\left\lceil\frac{(K(1-\frac{M}{N})-1)D}{K-1}\right\rceil}\cdots\right\rceil}\right\rceil$. Call such user vertex as $k_j$. Then,
$|\rho_j| \geq \left\lceil{\frac{(K(1-\frac{M}{N})-(j-1)}{K-(j-1)}\left\lceil {\cdots\left\lceil \frac{D(K(1-\frac{M}{N})-1)}{K-1}\right\rceil}\cdots\right\rceil}\right\rceil.$
Running over all $j$, and using (\ref{eqn1}), we therefore get the bound in the theorem.
\end{IEEEproof}
%
For a number of parameters we now compare (in Table \ref{tab2}) the above new bound on the number of transmissions (column 4 of Table \ref{tab2}) in an optimal scheme, with the lower bound given in \cite{cheng2017coded} (given in column 5, which holds for PDA based schemes), as well as the lower bound (column 6) based on the Ali-Niesen rate ($R^*\geq \frac{K(1-\frac{M}{N})}{1+\frac{MK}{N}}$) as shown in \cite{WTP}. The bound given in \cite{cheng2017coded} used in Table \ref{tab2} is as follows.   

\begin{theorem}\cite{cheng2017coded}
$R^{*}F \geq  \left\lceil{\frac{DK}{F}}\right\rceil +  \left\lceil{\frac{D-1}{F-1}\left\lceil{\frac{DK}{F}}\right\rceil}\right\rceil +\cdots\\ 
\hspace{-0.3cm}
\cdots +
\left\lceil{\frac{1}{\frac{FM}{N}+1}\left\lceil {\frac{2}{\frac{FM}{N}+2}\left\lceil\cdots\left\lceil{\frac{DK}{F}}\right\rceil\cdots\right\rceil}\right\rceil}\right\rceil.$
\end{theorem}

It can be seen that for many of the parameters, our bound is better than those in \cite{cheng2017coded},\cite{WTP}. Further, the last column of Table \ref{tab2} denotes the rate achieved by the scheme in Section \ref{ourscheme} in this work, for whichever parameters are applicable.  

\begin{table}[htbp]
\centering
\begin{tabular}{|c|c|c|c|c|c|c|}
\hline
$K$ & $F$ &$D$ &[this work]& \cite{cheng2017coded}  & \cite{WTP} & Scheme 
\\ 
&&&&&& [Sec \ref{ourscheme}]\\
&&& $R^{*}F \geq $& $ R^{*}F \geq $& $ R^{*}F \geq $ & $RF$
\\

\hline
 15 & 50 & 30 &71 & 54 &65 & NA\\

\hline
24 & 54 & 36 & 109 & 90 &96 & NA\\
\hline
15 & 20 &12 &30 & 31 &26 & NA\\
\hline
7 & 42 & 24 & 43 & 33 & 42& 56\\
\hline
15 & 210 & 168 & 637 & 444& 630 & 840\\
\hline
13 & 156 &108 &285 & 193 &280 & 468\\
\hline
\end{tabular}
\caption{\small For some values of $K,F,D$ , we compare the lower bound of this work with that of \cite{cheng2017coded}, \cite{WTP}. The last column gives the number of transmissions in the scheme constructed in this paper for whatever values are applicable. 
}
\label{tab2}
\end{table}
\section{A new projective geometry based scheme}
\label{ourscheme}
In this section we present a new coded caching scheme using projective geometries over finite fields. We first review some basic concepts.
\subsection{Review of projective geometries over finite fields \cite{hirschfeld1998projective}}
Let $k,q\in \mathbb{Z}^+$ such that $q$ is a prime power. Consider a $k$-dim (we use ``dim'' for dimensional) vector space $\fq^k$ over a finite field $\fq$. Consider an equivalence relation on $\fq^k\setminus \{\boldsymbol{0}\}$(where $\boldsymbol{0}$ represents the zero vector) whose equivalence classes are $1$-dim subspaces(without $\boldsymbol{0}
$) of $\fq^k$. The set of these equivalence classes is called the $(k-1)$-dim \textit{projective space} over $\fq$ and is denoted by $PG_q(k-1)$. For $m\in [k]$, let $PG_q(k-1,m-1)$ denote the set of all $m$-dim subspaces of $\fq^k$.
It is known that (Chapter $3$ in \cite{hirschfeld1998projective}) $|PG_q(k-1,m-1)|$ is equal to the \textit{q-binomial coefficient} $\gbinom{k}{m}$, where
$
\begin{bmatrix}k\\m\end{bmatrix}_q
=\frac{(q^k-1)\hdots(q^{k-m+1}-1)}{(q^m-1)\hdots(q-1)}.
$ 
The following result is known from \cite{hirschfeld1998projective}.
\begin{lemma}[Chapter 3 in \cite{hirschfeld1998projective}] \label{no of subspaces}
Consider a $k$-dim vector space $\mathbb{F}_q^k$. Let $1\leq r,s,l <k$. Then the number of $r$-dim subspaces intersecting a fixed $s$-dim subspace in a fixed $l$-dim subspace is $q^{(r-l)(s-l)}\gbinom{k-s}{r-l}$.

    
    



\end{lemma}

We now proceed to construct a caching line graph using projective geometry.
\subsection{A new caching line graph using projective geometry}
Consider $k,m,t \in \mathbb{Z}^+$ such that $m+t\leq k$. Let $W$ be a fixed $(t-1)$-dim subspace of the vector space $\fq^{k}$. 

Let 
\begin{align*}
    \mathbb{V} &\triangleq \{V \in PG_q(k-1,t-1): W\subseteq V \}.\\
    \mathbb{P} & \triangleq \{P \in PG_q(k-1,m+t-1): W\subseteq P\}.\\ 
    \mathbb{X} & \triangleq \left\{\{V_1,V_2,\cdots,V_{m+1}\}: \forall V_i \in \mathbb{V}, \sum\limits_{i=1}^{m+1}V_i \in \mathbb{P} \right\}.
\end{align*}
We first initialize $\mathcal{L}$ by its user-cliques. The user-cliques are indexed by $t$-dim subspaces in $\mathbb{V}$. For each $V\in \mathbb{V}$ create the vertices corresponding to the user-clique indexed by $V$ as
$C_V \triangleq \left\{(V,X) :X\in \mathbb{X}, V\nsubseteq \sum\limits_{V_i\in X}V_i\right\}$. Now, for each $X\in \mathbb{X}$ we construct the subfile clique of ${\C L}$ associated with $X$ as $C_{X}\triangleq \left\{(V,X): V\in \mathbb{V},V\nsubseteq \sum\limits_{V_i\in X}V_i \right\}$. By definition, these subfile cliques partition the set of vertices in $\cal L$ (the union of all the user cliques).  By invoking the notations from Section \ref{review} (Lemma \ref{linegraph}), we have $K=|\mathbb{V}|$ (number of user-cliques), 
and subpacketization $F=|\mathbb{X}|$ (the number of subfile cliques). We now find the values of $K$, and the size of the subfile cliques and the user cliques. 
\begin{lemma}
\label{K,c,D expressions}
\begin{align*}
K &= \gbinom{k-t+1}{1}. \\
|C_X| &=q^{m+1}\gbinom{k-m-t}{1} (\textit{for any }X\in {\mathbb{X}}). \\
|C_V| &=\gbinom{k-t}{m+1}\frac{q^{m+1}\prod\limits_{i=0}^{m}(q^{m+1}-q^{i})}{(q-1)^{m+1}(m+1)!} (\textit{for any } V\in{\mathbb{V}}).
\end{align*}    

\end{lemma}

\begin{IEEEproof}
$K=|\mathbb{V}|$ is the number of $t$-dim subspaces intersecting the fixed $(t-1)$-dim subspace $W$ in the fixed $(t-1)$-dim subspace $W$. By Lemma \ref{no of subspaces},
we have $ K=q^{(t-t+1)(t-1-t+1)}\gbinom{k-t+1}{t-t+1}=\gbinom{k-t+1}{1}$.

$|C_X|$ $($for any $X\in \mathbb{X})$ is the number of $t$-dim subspaces intersecting the fixed $(m+t)$-dim subspace $\sum\limits_{V_i\in X}V_i$ in the fixed $(t-1)$-dim subspace $W$. By Lemma \ref{no of subspaces},
we have $|C_X|=q^{(t-t+1)(m+t-t+1)}\gbinom{k-m-t}{t-t+1}
=q^{m+1}\gbinom{k-m-t}{1}$.

We now obtain $|C_V|$ for any $V\in\mathbb{V}$. To do this, we will first find the number (say $h$) of $(m+t)$-dim subspaces $P\in \mathbb{P}$ intersecting the fixed $t$-dim subspace $V$ in the fixed $(t-1)$-dim space $W$. By using Lemma \ref{no of subspaces}, we thus have $h=q^{(m+t-t+1)(t-t+1)}\gbinom{k-t}{m+t-t+1} = q^{m+1}\gbinom{k-t}{m+1}$. Now, we find the number(say $g$) of $X\in \mathbb{X}$ such that $\sum\limits_{V_i\in X}V_i =P$ for some $P\in \mathbb{P}$ . Then it follows that $|C_V|=hg$. 

Now, to find $g$, we first prove a few smaller claims.

\textit{Claim 1:} The number of one dimensional spaces $A$ such that $W\oplus A=V$  ($V$ being a fixed $t$-dimensional subspace, $\oplus$ representing direct sum) is $q^{t-1}$.~\\
\textit{Proof of Claim 1:} To see this, observe that to satisfy $W\oplus A=V$, we must have $A=span(\boldsymbol{a})$ for some $\boldsymbol{a}\in V\backslash W.$ Thus there are $q^t-q^{t-1}=q^{t-1}(q-1)$ choices for $\boldsymbol{a}$. However there are precisely $q-1$ vectors $\boldsymbol{a}$ whose span is the same one-dimensional subspace $A$. Hence we have that the number of one dimensional spaces $A$ such that $W\oplus A=V$ is $q^{t-1}.$

\textit{Claim 2: } Let $X=\{V_1,\hdots,V_{m+1}\}$ be some fixed element in ${\mathbb X}$. The number $N_1$ of $(m+1)$-sized sets $\{A_1,A_2,\cdots,A_{m+1}\}$ (where $A_i, \forall i\in [m+1]$ are $1$-dim subspaces)
such that $\{W\oplus A_1,W\oplus A_2,\cdots,W\oplus A_{m+1}\}=X$ is $q^{(t-1)(m+1)}.$ ~\\
\textit{Proof of Claim 2}: By Claim 1, the number of $A_i$ such that $W\oplus A_i=V_i$ is $q^{t-1}.$ As $A_i$s can be independently chosen to get the corresponding $V_i$s, we thus have that $N_1$ is $q^{(t-1)(m+1)}.$

\text{Claim 3:} The number $g'$ of $(m+1)$-sized sets $\{A_1,A_2,\cdots,A_{m+1}\}$ (where $A_i, \forall i\in [m+1]$ are $1$-dim subspaces)
such that $W\oplus A_1 \oplus A_2 \oplus \cdots A_{m+1}=P$ (for some fixed $P\in{\mathbb P}$) is $\dfrac{\prod\limits_{i=0}^{m}(q^{m+t}-q^{t-1+i})}{(q-1)^{m+1}(m+1)!}.$~\\

\textit{Proof of Claim 3:} Firstly, we note that there exists a  set $\{A_1,...,A_{m+1}\}$ of $1$-dim subspaces such that $W\oplus\bigoplus\limits_{i=1}^{m+1}A_i=P$, \textit{if and only if} there exists a (not necessarily unique) set of vectors $\boldsymbol{a_i}:i\in[m+1]$ such that $A_i=span(\{\boldsymbol{a_i}\})$ and $\boldsymbol{a_i}\in P\backslash(W\oplus\bigoplus\limits_{j=1}^{i-1}A_j)$, for all $i\in[m+1]$. We call such a set $\{\boldsymbol{a_i}:i=1,..,m+1\}$ as a \textit{generating set} of the set $\{A_i:i=1,..,m+1\}$.

Note that given a set of $1$-dim subspaces $\{A_i:i=1,..,m+1\}$, we can get a generating set $\{\boldsymbol{a_i}:i=1,..,m+1\}$ by choosing any $\boldsymbol{a_i}\in A_i\backslash \{\boldsymbol{0}\}$ for each $i\in[m+1]$. Now, suppose the set $\{\boldsymbol{a_i}:i\in[m+1]\}$ is a generating set for $\{A_i:i\in[m+1]\}$, then so is $\{c_i\boldsymbol{a_i}:i\in[m+1]\}$, for any $c_i\in{\mathbb F}_q\backslash \{0\}.$ Thus, the number of such distinct generating sets for any given set of $1$-dim subspaces $\{A_i:i\in[m+1]\}$ is $(q-1)^{m+1}$. 

Now the number of ways to choose an \textit{ordered} sets of vectors $\{\boldsymbol{a_i}:i\in[m+1]\}$ such that $\boldsymbol{a_i}\in P\backslash(W\oplus\bigoplus\limits_{j=1}^{i-1}A_j)$, for all $i\in[m+1]$, is $\prod\limits_{i=0}^{m}(q^{m+t}-q^{t-1+i})$. Thus, the number of (unordered) such generating sets $\{\boldsymbol{a_i}:i\in[m+1]\}$ is then $\frac{\prod\limits_{i=0}^{m}(q^{m+t}-q^{t-1+i})}{(m+1)!}$. By arguments in the previous paragraph, it can be seen that these unordered generating sets can be partitioned into groups of $(q-1)^{m+1}$, such that each such group includes precisely the set of all generating sets for a particular set of $1$-dim spaces $\{A_i:i\in[m+1]\}$ such that $W\oplus\bigoplus\limits_{i=1}^{m+1}A_i=P$. 

Thus the number $g'$ we are looking for is precisely $\dfrac{\prod\limits_{i=0}^{m}(q^{m+t}-q^{t-1+i})}{(q-1)^{m+1}(m+1)!}.$ This proves Claim 3. 

We now prove that the number $g$ of $X\in{\mathbb X}$ such that $\sum\limits_{V_i\in X}V_i=P$ for some $P\in{\mathbb P}$ is $\frac{\prod\limits_{i=0}^{m}(q^{m+1}-q^{i})}{(q-1)^{m+1}(m+1)!}.$ To see this, note that for each $X=\{V_1,...,V_{m+1}\}\in {\mathbb X}$ such that $\sum\limits_{V_i\in X}V_i=P$, there exists precisely $q^{(t-1)(m+1)}$ sets of $1$-dim subspaces $\{A_1,..,A_{m+1}\}$ such that $\{W\oplus A_i:i\in[m+1]\}=\{V_i:i\in[m+1]\}.$ By Claim 3, the total number of  sets of $1$-dim subspaces $\{A_1,..,A_{m+1}\}$ such that $W\oplus\bigoplus\limits_{i=1}^{m+1}A_i=P$ is $g'$, and these can be partitioned into groups each of size $q^{(t-1)(m+1)}$, such that for each set $\{A_1,..,A_{m+1}\}$ in any particular group, the set $X=\{V_i=W\oplus A_i:i\in[m+1]\}\in {\mathbb X}$ is the same. Thus we have that $g=\frac{g'}{q^{(t-1)(m+1)}}=\dfrac{\prod\limits_{i=0}^{m}(q^{m+1}-q^{i})}{(q-1)^{m+1}(m+1)!}.$ 

Finally, we see that the expression for $|C_V|=hg$ matches the lemma statement, which proves the lemma.





\end{IEEEproof}
Note that by Lemma \ref{K,c,D expressions}, we have the size of the subfile cliques of ${\cal L}$ as $|C_X|$ (for any $X\in{\mathbb X}$). We now show that $\Lbarsq$ has a clique cover with $d$-sized disjoint cliques for some $d$. Therefore ${\cal L}$ is in fact a $(c=|C_X|,d)$-caching line graph, giving raise to the main result in this section which is Theorem \ref{main result}. 
\subsection{Delivery Scheme from a clique cover of $\Lbarsq$}
We first describe a clique of $\Lbarsq$ and show that such equal-sized cliques partition $V(\mathcal{L})=V(\Lbarsq)$. This will suffice to show the delivery scheme as per Theorem \ref{cliquecoverlinegraph}. 

Let $\mathbb{Y}=\Big\{\{V_1,V_2,\cdots,V_{m+2}\}: V_i \in \mathbb{V}, \forall i\in [m+2]$ such that $ \sum\limits_{i=1}^{m+2}V_i \in PG_q(k-1,m+t)\Big\}$. We now present a clique of size $(m+2)$ in ${\Lbarsq}$.
\begin{lemma} \label{size of transmission clique}
Consider $Y=\{V_1,V_2,\cdots,V_{m+2}\}\in \mathbb{Y}$. Then
$C_Y=\left\{(V_i,Y\setminus V_i),\forall V_i\in Y\right\}\subset V(\Lbarsq)$ is a clique in $\overline{\mathcal{L}^2}$.
\end{lemma}
\begin{IEEEproof}
First note that $C_Y$ is well defined as $V_i\nsubseteq \sum\limits_{\substack{l=1 \\l\neq i}}^{m+2}V_l$ (otherwise $\sum\limits_{i=1}^{m+2}V_i$ will not be a $(m+t+1)$-dim space) and hence $(V_i,Y\setminus V_i)\in V({\C L})$. Consider two distinct vertices $(V_i,Y\setminus V_i),(V_j,Y\setminus V_j)\in C_Y$. It is clear that $V_i\neq V_j$ and $Y\setminus V_i \neq Y\setminus V_j$. By the construction of $C_Y$ we have $V_i \subseteq \sum\limits_{\substack{l=1 \\l\neq j}}^{m+2}V_l$ and $V_j \subseteq \sum\limits_{\substack{l=1 \\l\neq i}}^{m+2}V_l$. Therefore we have  $(V_i,Y\setminus V_j), (V_j,Y\setminus V_i) \notin V(\overline{\mathcal{L}^2})$. By invoking Lemma \ref{edge in compliment of square}, $\{(V_i,Y\setminus V_i),(V_j,Y\setminus V_j)\} \in E(\Lbarsq)$. Hence proved.
\end{IEEEproof}

Now we show that the cliques $\{C_Y : Y\in \mathbb{Y}\}$ partition $V({\Lbarsq})$.
\begin{lemma} \label{partition of L}
$\bigcup\limits_{Y\in \mathbb{Y}}C_Y=V(\mathcal{L})$,
where the union is a disjoint union.
\end{lemma}
\begin{IEEEproof}
Consider $Y, Y'\in \mathbb{Y}$ such that $Y\neq Y'$. By definition of $C_Y$, we have $C_Y \cap C_{Y'}=\phi$. Now consider an arbitrary vertex $(V_1,\{V_2,V_3,\cdots,V_{m+2}\}) \in V(\mathcal{
L})$. By the construction of $\mathcal{L}$,  $ \sum\limits_{i=1}^{m+2}V_i \in PG_q(k-1,m+t)$. Therefore $(V_1,\{V_2,V_3,\cdots,V_{m+2}\})$ lies in the clique, $ C_{\{V_1,V_2,\cdots,V_{m+2}\}} $ (defined as in Lemma \ref{size of transmission clique}). Hence proved.
\end{IEEEproof}

Finally we present our coded caching scheme using the caching line graph constructed above.

\begin{theorem}\label{main result}
The caching line graph $\mathcal{L}$ constructed above is a $ \left(c=q^{m+1}\gbinom{k-m-t}{1}, d=m+2 \right)$-caching line graph and defines a coded caching scheme with

$ K=\gbinom{k-t+1}{1}$, 

$ F=\gbinom{k-t+1}{m+1}\dfrac{\prod\limits_{i=0}^{m}(q^{m+1}-q^{i})}{(q-1)^{m+1}(m+1)!} $, 

$ \frac{M}{N}=1-\dfrac{q^{m+1}\gbinom{k-t}{m+1}}{\gbinom{k-t+1}{m+1}}$, and $ R=\dfrac{q^{m+1}\gbinom{k-m-t}{1}}{m+2}$.
\end{theorem}
\begin{IEEEproof}
From Lemma \ref{K,c,D expressions} and the notations in Lemma \ref{linegraph}, we get the expression of $K$ and $D=|C_V|.$
Further we see that the subfile cliques partition the vertices of ${\cal L}$ by definition and also $c=|C_X|$ for any $X\in \mathbb{X}$ (the size of each subfile clique). By Lemma
\ref{size of transmission clique} and Lemma \ref{partition of L}, the size of the cliques of  $\Lbarsq$ is $(m+2)$ and they partition the vertices. Hence  ${\C L}$ is a  $\small \left(c=q^{m+1}\gbinom{k-m-t}{1}, d=m+2 \right)$-caching line graph. 

Thus, we have by Theorem \ref{cliquecoverlinegraph},

\[F=\frac{KD}{c}=\gbinom{k-t+1}{m+1}  \dfrac{\prod\limits_{i=0}^{m}(q^{m+1}-q^{i})}{(q-1)^{m+1}(m+1)!}.\] 
\[\frac{M}{N}=1-\frac{c}{K}=1-\dfrac{q^{m+1}\gbinom{k-t}{m+1}}{\gbinom{k-t+1}{m+1}}.\]
\[\text{Since }  \frac{\gbinom{k-t+1}{1}}{\gbinom{k-m-t}{1}}= \frac{\gbinom{k-t+1}{m+1}}{\gbinom{k-t}{m+1}}.\]
\[R=\frac{c}{d}=\dfrac{q^{m+1}\gbinom{k-m-t}{1}}{m+2}.\]
This completes the proof.
\end{IEEEproof}

\section{Asymptotic analysis of the proposed scheme}
\label{asymptotics}
In this section, we analyse the behaviour of $R,F$ for the coded caching scheme proposed in Section \ref{ourscheme} as $1-\frac{M}{N}$ is lower bounded by a constant and $K\rightarrow \infty$. We show that $F=q^{O((log_qK)^2)}$, while $R=\Theta(\frac{K}{log_qK})$. Towards this end, we first give some bounds on  $q$-binomial coefficients. These can be easily derived, however a proof is available in \cite{PK}.
\begin{lemma}\cite{PK}
\label{approximations}
For non-negative integers $a,b,f$, for $q$ being some prime power, 
\begin{align*}
q^{(a-b)b}&\leq \gbinom{a}{b} \leq  q^{(a-b+1)b}\\
q^{(a-f-b-1)\delta} &\leq \frac{\gbinom{a}{b}}{\gbinom{a}{f}} \leq  q^{(a-f-b+1)\delta}, \\
\textit{where } \delta=|b-f|.
\end{align*}


\end{lemma}

We now proceed to analyse the asymptotics of the scheme. Throughout our analysis we assume $q$ is constant. Consider, 
\begin{align*}
\label{eqn34}
1-\frac{M}{N}=\frac{q^{m+1}(q^{k-t-m}-1)}{(q^{k-t+1}-1)} \geq q^{(m-k+t)}(q^{k-t-m}-1).
\end{align*}
To lower bound $1-\dfrac{M}{N}$ by a constant, let $k-m-t=
\alpha$, where $\alpha$ is a constant. Note that $\alpha \geq 0$ as $m+t \leq k$. We have $K=\gbinom{k-t+1}{1}$. We analyse our scheme as $(k-t)$ grows large. By Lemma \ref{approximations}, we have $q^{k-t}\leq K \leq q^{k-t+1}$. Hence we have
\begin{equation}
\label{eqn35}
    {\log_{q}{K}}-1 \leq (k-t)\leq \log_{q}{K}.
\end{equation}
The rate expression in Theorem \ref{main result} can be written as 
$R=\dfrac{K(1-\frac{M}{N})}{m+2} = \dfrac{K(1-\frac{M}{N})}{k-t-\alpha+2}$.

Therefore by using (\ref{eqn35}) we have
$\frac{K(1-\frac{M}{N})}{\log_q{K}-\alpha+2} \leq R \leq \frac{K(1-\frac{M}{N})}{\log_q{K}-1-\alpha+2},$

Consider, $R \leq \frac{K(1-\frac{M}{N})}{\log_q{K}-1-\alpha+2} = \frac{K(1-\frac{M}{N})}{\log_q{K}} (1-\frac{\alpha-1}{\log_q{K}})^{-1}$. Now  by Taylor's series expansion we have $1\leq (1-\frac{\alpha-1}{\log_q{K}})^{-1}\leq 2$ as $K\rightarrow \infty$. Therefore we have $R \leq \frac{2K(1-\frac{M}{N})}{\log_q{K}}$. Hence $R = O(\frac{K}{\log_q{K}})$.

Consider, $R \geq \frac{K(1-\frac{M}{N})}{\log_q{K}-\alpha+2} = \frac{K(1-\frac{M}{N})}{\log_q{K}} (1-\frac{\alpha-2}{\log_q{K}})^{-1}$. Now if $\alpha \geq 2$, by Taylor's series expansion we have $1\leq (1-\frac{\alpha-2}{\log{K}})^{-1}\leq 2$ as $K\rightarrow \infty$. Therefore we have $R \geq \frac{K}{\log_q{K}}$. If $ \alpha \in \{0,1\}$, then $R \geq \frac{K(1-\frac{M}{N})}{\log_{q}{K}-\alpha+2} \geq \frac{K(1-\frac{M}{N})}{2\log_{q}{K}}$. Hence $R = \Omega(\frac{K}{\log_q{K}})$. Therefore $R=\Theta \left(  \dfrac{K}{\log_q{K}}\right)$. 

We now obtain the asymptotics for subpacketization $F$. By Lemma \ref{approximations} and from the expression of $F$ in Theorem \ref{main result}, we have





\begin{align*}
    F&=K.\frac{\gbinom{k-t+1}{m+1}}{K}.\frac{\prod\limits_{i=0}^{m}(q^{m+1}-q^{i})}{(q-1)^{m+1}(m+1)!}\\
    &\leq \frac{Kq^{(k-t-m)m}\prod\limits_{i=0}^{m}q^{i}(q^{m+1-i}-1)}{(q-1)^{m+1}(m+1)!}\\
    &=\frac{Kq^{\alpha m}}{(m+1)!}\prod\limits_{i=0}^{m} q^{i}\gbinom{m+1-i}{1}.
    \end{align*}
Once again, by applying Lemma \ref{approximations} to the above expression, we have,
$F \leq \frac{Kq^{\alpha m}}{(m+1)!}\prod\limits_{i=0}^{m} q^{i} q^{m+1-i}=\frac{q^{\log_q{K}+\alpha m+(m+1)^2}}{(m+1)!}. 
$ As $m+t\leq k$, thus $q^{m}\leq q^{k-t}\leq K$ (by Lemma \ref{approximations}). Hence $m\leq \log_q{K}$. 
Also by (\ref{eqn35}) we have, ${\log_{q}{K}}-1 \leq (m+\alpha)$, which can be written as $\frac{1}{(m+1)!} \leq \frac{1}{\lfloor{\log_{q}{K}}-\alpha \rfloor !}$. Using Stirling's approximation for $y!$ as $\sqrt{2\pi x}(\frac{y}{e})^x$ for large $y$, and after some simplifications we see that $F=q^{O\left((\log_{q}{K})^2\right)}$.

\begin{table}
\begin{tabular}{|c|c|c|c|c|c|c|c|}
\hline
$(k,m,t,q)$ & $(m',q')$ & & & & & & \\ 
$K_1$ & $ K_2$ & $U_1$ & $U_2$ &$F_1$ &  $F_2$ &$\gamma_1$& $\gamma_2$\\
&\cite{YCTCPDA} & & \cite{YCTCPDA} & &\cite{YCTCPDA} & & \cite{YCTCPDA}\\
\hline
$(10,2,2,2)$ & $(6,73)$ &&&&&&\\
511 & 511 & 0.98 & 0.98 & $10^{7}$ & $10^{11}$ & 4 & 7
\\
\hline
$(9,3,2,2)$ & $(14,17)$ &&&&&&\\
255 & 255 & 0.94 & 0.94 & $10^{8}$ & $10^{17}$ & 5 & 15
\\
\hline

$(8,3,2,2)$ & $(13,9)$ &&&&&&\\
127 & 126 & 0.88 & 0.88 & $10^{6}$ & $10^{12}$ & 5 & 14
\\
\hline

$(9,4,3,2)$ & $(31,4)$ &&&&&&\\
127 & 128 & 0.76 & 0.75 & $10^{8}$ & $10^{18}$ & 6 & 32
\\
\hline

$(7,3,2,2)$ &$(15,4)$ &&&&&&\\
63 & 64& 0.76 & 0.75 & $ 10^5$ & $10^9 $ & 5 &16\\
\hline

$(7,3,3,3)$ & (39,3) &&&&&&\\
121 & 120 & 0.67 & 0.66 & $10^6$ & $10^{18} $ & 5 & 40\\
\hline
$(6,3,2,2)$ & $(14,2)$ &&&&&&\\
31 & 30 & 0.51 & 0.50 & $10^4$ & $10^4 $ & 5& 15\\
\hline
\end{tabular}
\caption{For some specific values of $K,U=1-\frac{M}{N}$, we compare the results of \cite{YCTCPDA} with this work.}
\label{tab3}
\end{table}

Finally in Table \ref{tab3}, we compare the scheme in Theorem \ref{main result} with the scheme in \cite{YCTCPDA} for some choices of $K,U=1-\frac{M}{N},F$ and $\gamma$ (the global caching gain, i.e., $\frac{K(1-\frac{M}{N})}{R}$, where $R$ is the rate achieved by the scheme). We label the parameters of our scheme in Theorem \ref{main result} as $K_1,U_1,F_1,\gamma_1$ where $\gamma_1=d$. The parameters of the scheme presented in \cite{YCTCPDA} are $K_2=q'(m'+1),U_2=1-\frac{1}{q'},F_2=(q')^{(m')},\gamma_2=\frac{K(1-\frac{M}{N})}{q'-1}$ where $q',m'\in \mathbb{Z}^+$. The first column lists $(k,m,t,q),K_1$ according to the Theorem \ref{main result}. The second column lists $(m',q'),K_2$ parameters of the scheme in \cite{YCTCPDA}. Since we can not match the parameters from this work and \cite{YCTCPDA} exactly, we choose approximately equal values. We see from the table that our scheme performs much better than \cite{YCTCPDA} 
in terms of the subpacketization, but pays a price in terms of the rate.

\bibliographystyle{IEEEtran}
\bibliography{IEEEabrv,cite.bbl}

\end{document}